\documentclass[pra,onecolumn,floatfix,a4paper,superscriptaddress]{revtex4}
\usepackage{bm,color,graphicx,amsmath,txfonts}

\usepackage[colorlinks, citecolor=blue,linkcolor=blue]{hyperref}

\begin{document}	
\title{ Enhancement of mirror-mirror entanglement with intracavity squeezed light and squeezed-vacuum injection}	
\author{Noura Chabar}
\affiliation{LPTHE-Department of Physics, Faculty of sciences, Ibnou Zohr University, Agadir, Morocco}	
\author{M'bark Amghar}
\affiliation{LPTHE-Department of Physics, Faculty of sciences, Ibnou Zohr University, Agadir, Morocco}	
\author{Mohamed Amazioug} \thanks{amazioug@gmail.com}
\affiliation{LPTHE-Department of Physics, Faculty of sciences, Ibnou Zohr University, Agadir, Morocco}	
\author{Mostafa Nassik}
\affiliation{LPTHE-Department of Physics, Faculty of sciences, Ibnou Zohr University, Agadir, Morocco}
\begin{abstract}		
In this manuscript, we investigate the enhancement of the transfer of quantum correlations from squeezed light to movable mirrors within an optomechanical system. This enhancement was achieved via the injection of squeezed light in the cavities and via intracavity squeezed light. We quantify the entanglement between mechanical oscillators via logarithmic negativity. We demonstrate that entanglement is influenced by various factors, including the gain of the parametric amplifier, the squeezing parameter characterizing the squeezed light, the rate of the phonon tunneling process, the coupling strength of the photon hopping process and the bath temperature of the mechanical oscillators. We have shown that entanglement can be improved by a convenient choice of coupling strength in the case of the photon hopping process, as well as for specified values of the gain of the parametric amplifier.\\
	
\vspace{0.25cm}\textbf{Keywords}: Cavity Optomechanics; Photon hopping Process ; Parametric amplifier; Phonon tunneling process; Quantum correlations; Logarithmic negativity; Entanglement.		
\end{abstract}	
\date{\today}
\maketitle	
\section{Introduction}
The cavity optomechanical system offers a promising platform for investigating the interaction between light and mechanical oscillators \cite{1,2}. This interaction arises due to the radiation pressure acting between the mobile mirror and the cavity field within the cavity \cite{3, amziougIJQI19, amziougChin19,1997,amziougsr23}. During the past two decades, cavity optomechanical systems have garnered significant interest, particularly in the realm of quantum information processing. They have been explored for various fascinating quantum tasks, such as achieving ground-state cooling of mechanical modes \cite{4,7}, creating mechanical quantum superpositions \cite{8}, realizing entanglement between mechanical and optical modes in steady state \cite{9,11}, quantum measurement's precision \cite{12,15} and gravitational-wave detectors \cite{16,18}. The exchange of quantum correlations among states within optomechanical systems holds great significance \cite{oo,imk,eleuch}. This importance stems primarily from the pivotal role that entanglement (non-separability) plays in enabling and augmenting quantum information processing, such as in the case of quantum teleportation \cite{ol}. Recently, there has been a particular emphasis on improving quantum correlations in stable states among mechanical modes \cite{30} and on investigating the mechanical entanglement of two Fabry-P\'erot cavities, where they are interconnected through the photon hopping  and each of these cavities is equipped with an internal  parametric amplifier \cite{r,p}. The suggested scheme  in our study for achieving this enhancement between the two moving mirrors, relies on employing a degenerate parametric amplifier and the injection of squeezed light within a double-cavity optomechanical system, in the presence of the photon hopping process and the phonon tunneling process. It is worth noting that the photon hopping process has been demonstrated to be a degrading factor for quantum correlations in optomechanical systems of this kind \cite{hmj,mm}. The interaction between the system and its surrounding environment leads to the dissipation of quantum correlations, and under specific conditions, this phenomenon can manifest suddenly. This is referred to as entanglement sudden death (ESD) \cite{SOL4}. Additionally, it has been shown that in certain specific situations, entanglement can be spontaneously created. This phenomenon is commonly referred to entanglement sudden birth (ESB) \cite{sd}.
This paper delves into the investigation of a quantum optomechanical setup, which involves two Fabry-P\'erot cavities. Each  cavity contains a parametric amplifier (PA) to enhance the entanglement of the mechanical components by generating squeezed light \cite{48}, we created two couplings in the system: the first between the optical modes through a photon hopping process, and the second between the mechanical resonators through the phonon tunneling process. Additionally, coherent sources and squeezed light are used to pump the system. The generated entanglement is quantified by the logarithmic negativity. \\
The paper's structure is outlined as follows: In Section II, we outline the model under consideration, give the expression of the Hamiltonian and the corresponding quantum Langevin equations for mechanical and optical modes. In section III, we introduced the EPR operators to obtain the covariance matrix in the steady state. In section IV, we provide the explicit expression of logarithmic negativity used to quantify mechanical entanglement. Section V delves into the evolution of this measure for mechanical modes, considering various influencing factors. Ultimately, section VI   presents the concluding remarks of the paper. 	
\section{Model and dynamical equations}
In this study we discuss the behavior of quantum correlations between two mechanical oscillators in an hybrid optomachanical system, where we have two spatially separated cavities, each of those cavities are consists of a movable mirror $(M_{i},i=1,2)$ and fixed mirror $(FM_{i},i=1,2) $. The cavities are driven by coherent laser sources  and  pumped by squeezed light, inside each of them we have a parametric amplifier  to generate squeezed light  \cite{asjad}, and whose  presence promotes optomechanical cooling  \cite{asjad}.
\begin{figure}
\begin{center}
\fbox{\includegraphics[width=0.45\textwidth]{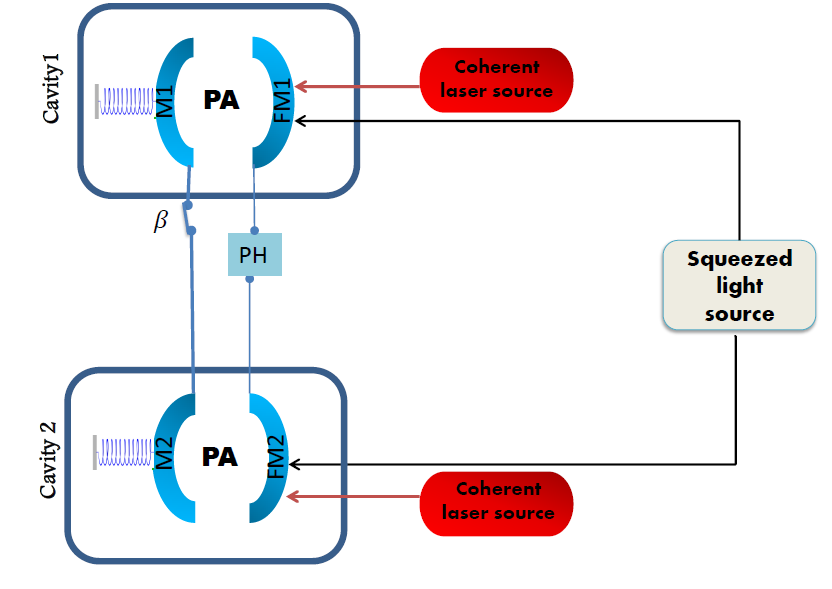}{(a)}}
\fbox{\includegraphics[width=0.43\textwidth]{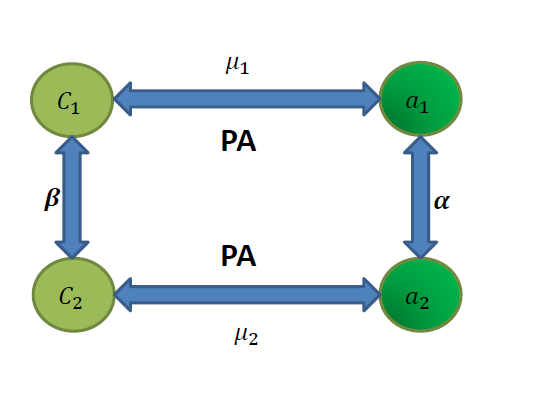}{(b)}}
\caption{(a) Schematics of two optomechanical cavities spatially separated. Both cavities are driven by coherent laser sources. Optical modes $c_{i}$ are coupled by a photon hopping process (PH) and mechanical modes ($a_{j}$) are coupled by a phonon tunneling process. (b) The various coupling strengths in the system, $\beta$ is the coupling rate of phonon tunneling process, $\alpha$ is the coupling force of photon hopping (PH) and $\mu_{j}$ representes the coupling between the jth mechanical mode and the intensity of optical mode, each cavity has inside a parametric amplifier.}
\label{fig:LENEV}
\end{center}
\end{figure}  
The Hamiltonian describing the system, writes as 
	\begin{equation}
		H= H_{Opm}+H_{Mm}+H_{In}+H_{Las}+H_{PA} +H_{\alpha}+H_{\beta}   
	\end{equation}
where 
$H_{Mm}= \hbar\sum_{j=1}^{2} \omega_{{M}_{j}}a_{j}^{+}a_{j}
$ and 	$H_{Opm}=	-\hbar\sum_{j=1}^{2}\Delta _{j}c_{j}^{+}c_{j}$ are the energy of  mechanical and  optical modes, respectively and $\Delta_j=\omega_{\ell_j}-\omega_{c_j}$ is the input-cavity detuning, $\omega _{c_{j}}$ are the  frequencies inside each cavity, the parameters $\omega _{M_{j}}$ are the frequencies of movable mirror and $ \omega_{l_{j}}$ are the frequencies of the jth input field. The operators $c_{j}(c_{j}^{+})$ and $a_{j}(a_{j}^{+})$ representes the annihilation (creation) operators of the jth cavity optical modes and  mechanical modes respectively, they satisfy the following canonical commutation $\left[a_j, a_j^{+}\right]=1; \left[c_j,c_j^{+}\right]=1$ $(j=1,2)$. $H_{In}= -\hbar \sum_{j=1}^{2}\mu_{j}c_{j}^{+}c_{j}(a_{j}^{+}+a_{j})$ is the energy of optomechanical coupling via radiation pressure where $ \mu _{j}=\frac{\omega_{c_j}}{L_j} \sqrt{\frac{\hbar}{m_j \omega_{M_j}}}$ representes the coupling between the jth mechanical mode and the intensity of optical mode \cite{1}, $L_j$ being the jth cavity length and $ m_{j}$ the jth mass of the movable mirror \cite{1}. $H_{Las}=\hbar \sum_{j=1}^{2}(c_{j}^{+}\upsilon_{j}\text{e}^{i\varphi_{j}}+c_{j}\upsilon_{j}\text{e}^{-i\varphi_{j}} )$ 
describes the energy of optical driving, with $ \varphi_{j}$ and $ \upsilon _{j}=\sqrt{\frac{2 \Gamma_j p_j}{\hbar \omega_{l_j}}}$ $(j=1,2)$ are, respectively,  the phase and the amplitude of the input field where $\Gamma_j$ is the jth cavity damping rate and  $ p_j$ is the drive pump power of the jth laser. The terms  $H_{PA}= i\hbar  \sum_{j=1}^{2}\lambda_{j}(\text{e}^{i \theta} c_{j}^{{+^{2}}}\text{e}^{-2i \omega_{ M_{j}}t}-\text{e}^{-i \theta} c_{j}^{{^{2}}}\text{e}^{2i \omega_{Mj}t})$ and $H_{\alpha}= -\hbar \alpha (c_{1}^{+}c_{2}+c_{2}^{+}c_{1})$  express respectively,  the coupling between the optical modes inside the cavities, parametric amplifier and the photon hopping process  where  $\alpha$ is the coupling strength of the phonon hopping process, $\lambda_{j}$ and  $ \theta_{j}$  are, respectively, the gain and the phase of the jth pump field driving the parametric amplifier which is related to the pump driving the PA. By doing this, the pump field driving the PA at frequency $ 2(\omega_{ M_{j}}+\omega_{ l_{j}}) $ interacts with the second-order nonlinear optical crystal, i.e., the signal and the idler have identical frequency $ (\omega_{ M_{j}}+\omega_{ l_{j}}) $ \cite{ideler,30} . The last term $H_{\beta}=  - \hbar  \beta (a_{1}^{+}a_{2}+ a_{2}^{+}a_{1})$ is the energy of coupling between movable mirrors via photon tunneling process and $\beta$ is the coupling force between the moving mirrors trough a phonon tunneling process. The dynamics of mechanical and the optical modes satisfy the  following nonlinear Langevin equation, given by
\begin{equation}
\begin{gathered}
\dot{a_1}=-\left(\frac{\gamma_1}{2}+\mathrm{i} \omega_{M_1}\right) a_1+\mathrm{i} \mu_1 c_1^{+} c_1+\mathrm{i} \beta a _{2}+\sqrt{\gamma_1} a_1^{i n}
\end{gathered}
\end{equation}
\begin{equation}
\dot{a_2}=-\left(\frac{\gamma_2}{2}+\mathrm{i} \omega_{M_2}\right) a_2+\mathrm{i} \mu_2 c_2^{+} c_2+\mathrm{i} \beta a _{1}+\sqrt{\gamma_2} a_2^{i n}
\end{equation}
\begin{equation}
\dot{c_1}=-\left(\frac{\Gamma_1}{2}-\mathrm{i} \Delta_1\right) c_1+\mathrm{i} \mu_1 c_1\left(a_1^{+}+a_1\right)-\mathrm{i} \vartheta_1 \mathrm{e}^{\mathrm{i} \varphi_1}+2 \lambda_1 \mathrm{e}^{\mathrm{i} \theta} c_1^{+} \mathrm{e}^{-\mathrm{i} 2 \omega_{M_1} t}+\mathrm{i} \alpha c_2+\sqrt{\Gamma_1} c_1^{i n}
\end{equation}
\begin{equation}
\dot{c_2}=-\left(\frac{\Gamma_2}{2}-\mathrm{i} \Delta_2\right) c_2+ \mathrm{i} \mu_2 c_2\left(a_2^{+}+a_2\right)-\mathrm{i} \vartheta_2 \mathrm{e}^{\mathrm{i} \varphi_2}+2 \lambda_2 \mathrm{e}^{\mathrm{i} \theta} c_2^{+} \mathrm{e}^{-\mathrm{i} 2 \omega_{M_2} t}+\mathrm{i}\alpha c_{1}+\sqrt{\Gamma_2} c_2^{i n}
\end{equation}
where $ \gamma_{j}$   is the dissipation rate of the jth movable mirror. The squeezed vacuum operators $c_j^{i n}$ and $ c_j^{{+}{i n}}$ verify the following non-zero correlations  relations \cite{41}
\begin{equation}
\begin{aligned}
& \left\langle c_j^{i n }(t)c_j^{  {i n } {\dagger}}(t')\right\rangle = (\mathcal{R}+1)\delta (t-t') \\
&\left\langle c_j^{{i n } {\dagger}}     (t)c_j^{i n }(t')\right\rangle = \mathcal{R} \delta (t-t')\\ 
&\left\langle c_j^{i n }(t)c_{j'}^{i n }(t')\right\rangle =\mathcal{V} \mathrm{e}^{-\mathrm{i} \omega _{M}(t+t')}\delta (t-t') \quad ;\quad  \mathrm{j}  \neq \mathrm{j'} \\
&\left\langle c_j^{{i n } {\dagger}}      (t)c_{j'}^{{i n } {\dagger}}     (t')\right\rangle =\mathcal{V} \mathrm{e}^{\mathrm{j} \omega _{M}(t+t')} \delta (t-t') \quad; \quad  \mathrm{j}  \neq \mathrm{j'}
\end{aligned}
\end{equation}
where $\mathcal{R}= \sinh^{2}(r) $, $ \mathcal{V}= \sinh(r)\cosh(r)$ and $r$ stands for the squeezing parameter characterizing the squeezed light. The terms 
$a_j^{i n}$ represent the jth noise operators describing the coupling between the movable mirror and its own environment, and it can be presumed that the mechanical baths are Markovian, $a_j^{i n}$ and $a_j^{{i n}{\dagger}}$ has zero-mean value and we have \cite{b,c}
\begin{equation}
\begin{aligned}
&\left\langle a_{j}^{in}(t) a_{j}^{{in}{\dagger}}\left(t^{\prime}\right)\right\rangle=\left(n_{{th}_{j}}+1\right) \delta\left(t-t^{\prime}\right).\\
&\left\langle a_{j}^{{in} {\dagger}}(t) a_{j}^{{in}{\dagger}}(t^{\prime}) \right\rangle=n_{{th}_{j}} \delta\left(t-t^{\prime}\right)
\end{aligned}
\end{equation}
where $n_{{th}_{j}}=\left[\exp \left(\hbar \omega_{M_j} /\left(k_B T_j\right)\right)-1\right]^{-1}
$ is the photon number in the jth cavity, $k_{B}$ is the Boltzmann constant. The nonlinear quantum Langevin equations are in general non-solvable analytically, for that we use the following linearisation scheme 
$\mathcal{T}_{j}=\left\langle \mathcal{T} _{j}\right\rangle + \delta \mathcal{T}_{j} $ where  $\mathcal{T}_{j}$ replace the two operators $ a_{j}$ and $c_{j}$, $\left\langle \mathcal{T} _{j}\right\rangle $ are the mean value in the steady state and $\delta \mathcal{T}_{j}$ are the operators of fluctuation \cite{eleuch}.  The values at steady-state are given by
\begin{equation*}
\left\langle a _{1}\right\rangle = \frac{\mathrm{i}\mu_{1}\mathcal{I}_{2}|\left\langle c_{1}\right\rangle  |^{2}-\beta \mu _{2}|\left\langle c_{2}\right\rangle |^{2}}{\mathcal{I}_{2}\mathcal{I}_{1}+\beta ^{2}}
\quad ;  \quad  	\left\langle a _{2}\right\rangle = \frac{\mathrm{i}\mu_{2}\mathcal{I}_{1}|\left\langle c_{2}\right\rangle  |^{2}-\beta \mu _{1}|\left\langle c_{1}\right\rangle |^{2}}{\mathcal{I}_{1}\mathcal{I}_{2}+\beta ^{2}}
\end{equation*}
\begin{equation*}
\left\langle c _{1}\right\rangle = \frac{\mathrm{i}\upsilon_{1}\mathcal{B}_{2}\mathrm{e}^{\mathrm{i}\varphi_{1}}+\alpha \upsilon _{2}\mathrm{e}^{\mathrm{i}\varphi_2}}{\mathcal{B}_{2}\mathcal{B}_{1}+\alpha ^{2}}
\quad  ;   \quad  	\left\langle c _{2}\right\rangle = \frac{\mathrm{i}\upsilon_{2}\mathcal{B}_{1}\mathrm{e}^{\mathrm{i}\varphi_{2}}+\alpha \upsilon _{1}\mathrm{e}^{\mathrm{i}\varphi_1}}{\mathcal{B}_{1}\mathcal{B}_{2}+\alpha ^{2}}
\end{equation*}
With
$ \mathcal{I}_{j} = \mathrm{i} \omega_{M_ j}+ \frac{\gamma_{j}}{2} $, $ \mathcal{B}_{j}= - \frac{\Gamma_{j}}{2}+ \mathrm{i}\Delta_j^{\prime}  $ ($j=1,2$) and  $
\Delta_j^{\prime}=\Delta_j+\mu_j\left( \left\langle a  _{j}^{+}\right\rangle+ \left\langle a _{j}\right\rangle\right)
$. We consider that the two cavities are identical, with equal temperature 
$T_1=T_2=T  \left(n_{t h_1}=n_{t h_2}=n_{t h}\right)$, and then $ \lambda_1=\lambda_2=\lambda$, $ \alpha_1=\alpha_2 = \alpha$, $ m_1=m_2= m$, 
$\omega_{\ell_1}=\omega_{\ell_2}=\omega_{\ell}$, $\omega_{c_1}=\omega_{c_2}=\omega_c$, $ \omega_{M_1}=\omega_{M_2}=\omega_M$, $ \Gamma_1=\Gamma_2=\Gamma$, $\mathcal{B}_{1}= \mathcal{B}_{2}=\mathcal{B}$, $ \mathcal{I}_{1}= \mathcal{I}_{2}= \mathcal{I} $  and $\gamma_1=\gamma_2=\gamma$.	We noted that the many-photon optomechanical coupling within the jth cavity is defined as
$\mathcal{J}= \mu |\left\langle c \right\rangle|= \sqrt{\frac{2 \omega _{c}^{2}\Gamma P}{L^{2} m \omega_{M} \omega_{l}[ (\omega _M+ \alpha)^{2}  + ( \frac{\Gamma^{2}}{4})  ]}} $
\cite{40}. To describe any given phase of the coherent drives, we employ the phase of the input laser as $\varphi = - \arctan \left[ \frac{\Delta^{'}+\alpha} \Gamma\right] $ so that  $\left\langle c \right\rangle = \mathrm{i|\left\langle c \right\rangle|}$ \cite{48}.
In result the  linear  Langevin equations is given by the following equations, in the limit $ |\left\langle c \right \rangle |\gg 1 $ as 
\begin{equation}
\label{6}
{\delta} \dot a_{j}=-\left(\mathrm{i} \omega_{M}+\frac{\gamma}{2}\right) \delta a_{j}+\mathcal{J}\left(\delta c_{j}-\delta c_{j}^{+}\right)+\mathrm{i}\beta   \delta a_{n} +\sqrt{\gamma_{j}} a_{j}^{in} \quad ; \quad   \mathrm{j}\neq n
\end{equation}
\begin{equation}
\label{7}
{\delta} \dot c_{j}=-\left(\frac{\Gamma}{2}-\mathrm{i} \Delta^{\prime}\right) \delta c_{j}-\mathcal{J} \left(\delta a_{j}^{+}+\delta a_{j}\right)+ \mathrm{i}\alpha \delta c_{n}+2 \lambda \mathrm{e}^{\mathrm{i} \theta} \delta c_j^{+} \mathrm{e}^{-\mathrm{i} 2 \omega_{M} t} +\sqrt{\Gamma } c_{j}^{in}  \quad ; \quad \mathrm{j}\neq n
\end{equation}
with $
\Delta^{\prime}=\Delta+\mu \left( \left\langle a  _{j}^{+}\right\rangle+ \left\langle a _{j}\right\rangle\right)
$ is the effective cavity detuning and which depends on the displacement of the mirrors resulting  from  the radiation pressure force. To consider slow fluctuations, we introduce the following transformations: $	\delta c_{j}(t)=\delta\tilde{c}_{j}(t) ~e^{i \Delta^{\prime} t} and ~~~
\delta a_{j}(t)=\delta  \tilde{a}_{j}(t)~ e^{-i \omega_{M}t} $, for the noise operators we have:      
$ \tilde{c}_j^{i n} \rightarrow \mathrm{e}^{-\mathrm{i}  \Delta^{\prime} t} c_j^{i n}$ and $ \tilde{a}_j^{i n} \rightarrow \mathrm{e}^{\mathrm{i}\omega_{{M}} {t}} a_j^{i n} $.   
Omitting  the rapidly oscillating terms at $ \pm 2 \omega_{M}$, and assuming that the cavities are driven in the red sideband ($\Delta ^{{\prime}}= -\omega _{M}$). In  the rotational wave approximation (RWA), i.e. in the limit where the mechanical frequency $\omega _{M}$ is greater than the cavity decay rate   ($\omega _{M}\gg  \Gamma $) \cite{40}, the equations  (\ref{6}) and  (\ref{7}) became as follows
\begin{equation}   
\label{8}
\delta \dot{\tilde{a}}=-\frac{\gamma}{2} \delta \tilde{a}_{j}+\mathcal{J} \delta{\tilde{c}}_{j} +\mathrm{i}\beta\delta {\tilde{a}}_{n} +\sqrt{\gamma} \tilde{a}_{j}^{in}  \quad ; \quad   j\neq n
\end{equation}
\begin{equation}
\label{9}
\delta \dot{\tilde{c}}_{j}=-\frac{\Gamma}{2} \delta \tilde{c}_{j}-\mathcal{J} \delta {\tilde{a}}_{j}+\mathrm{i}\alpha \delta{\tilde{c}}_{n} +2\lambda (  \cos {\theta} + \mathrm{i} \sin {\theta} )\delta \tilde{c}_{j}^{+} +\sqrt{\Gamma} \tilde{c}_{j}^{in} \quad ;  \quad  j\neq n
\end{equation}
\section{The steady state }
The following EPR operators for the two mechanical and optical modes are introduced to obtain the covariance matrix
\begin{equation*}
\delta \tilde{Q}_{a_{j}} =\frac{\delta \tilde{a}_{j}^{+}+\delta \tilde{a}_{j}}{\sqrt{2}}~,\quad
\delta \tilde{P}_{a_{j}}=\frac{\delta \tilde{a}_{j}-\delta \tilde{a}_{j}^{+}}{i \sqrt{2}}~ ,  
\quad
\delta\tilde{Q}_{c j} =\frac{\delta\tilde{c}_{j}^{+}+\delta\tilde{c}_{j}}{\sqrt{2}}~,\quad
\delta\tilde{P}_{c_{j}} =\frac{	\delta\tilde{c}_{j}-	\delta\tilde{c}_{j}^{+}}{i \sqrt{2}}~,~ j = 1,2
\end{equation*} 
So its possible to rewrite the two equations (\ref{8}) and (\ref{9}) as follows
\begin{equation}
\label{etoile}
\delta \dot{\tilde{Q}}_{a j} =-\frac{\gamma}{2} \delta \tilde{Q}_{a j}+\mathcal{J} \delta \tilde{Q}_{c j}-\beta \delta \tilde{P}_{a_ {n}}+\sqrt{\gamma} \tilde{Q}_{a j}^{\text {in }} \quad ; \quad    j\neq n
\end{equation}
\begin{equation*}
\delta \dot{\tilde{P}}_{a j} =-\frac{\gamma}{2} \delta \tilde{P}_{a j}+\mathcal{J} \delta   \tilde{P}_{c j}+\beta \delta \tilde{Q}_{a_ {n}}+\sqrt{\gamma} \tilde{P}_{a j}^{\text {in }} \quad ; \quad  j\neq n
\end{equation*}
\begin{equation*}
\delta \dot{\tilde{Q}}_{c_{j}} =-\frac{\Gamma}{2} \delta \tilde{Q}_{c_{j}}-\mathcal{J} \delta \tilde{Q}_{a j}-\alpha \delta \tilde{P}_{c_ {n}}+2 \lambda( \cos{\theta}\delta \tilde{Q}_{c j}+ \sin{\theta}\delta \tilde{P}_{c j})
+\sqrt{\Gamma} \tilde{Q}_{c_{j}}^{\text {in }}  \quad ; \quad  j\neq n
\end{equation*}
\begin{equation}
\label{etoile double}
\delta \dot{\tilde{P}}_{c_ {j}} =-\frac{\Gamma}{2} \delta \tilde{P}_{c_{j}}-\mathcal{J} \delta \tilde{P}_{a j}+\alpha \delta \tilde{Q}_{c_ {n}}-2 \lambda( \cos{\theta}\delta \tilde{P}_{c j}- \sin{\theta}\delta \tilde{Q}_{c j})
+\sqrt{\Gamma} \tilde{P}_{c_{j}}^{\text {in }}  \quad ; \quad   j\neq n 
\end{equation}
where
\begin{equation}
\begin{array}{ll}
\tilde{Q}_{c_j}^{\mathrm{in}}=\frac{\tilde{c}_j^{\mathrm{in^{+}} }+\tilde{c}_j^{\mathrm{in}}}{\sqrt{2}}, \quad \tilde{P}_{c_j}^{\mathrm{in}}=\frac{\tilde{c}_j^{\mathrm{in}}-\tilde{c}_j^{\mathrm{in^{+}}}}{\mathrm{i} \sqrt{2}},\quad
\tilde{Q}_{a_j}^{\mathrm{in}}=\frac{\tilde{a}_j^{\mathrm{in^{+}}}+\tilde{a}_j^{\mathrm{in}}}{\sqrt{2}}, \quad \tilde{P}_{a_j}^{\mathrm{in}}=\frac{\tilde{a}_j^{\mathrm{in}}-\tilde{a}_j^{\mathrm{in^{+}}}}{\mathrm{i} \sqrt{2}},
\end{array}
\end{equation}
One could write these equations in the following matrix form
\begin{equation}
\dot{\mathcal{ Z}}(t)=\mathcal{Q} \mathcal{Z}(t)+\mathcal{Y}(t) 
\end{equation}
with $\mathcal{Z}(t)=\left(\delta \widetilde{Q}_{a_1}, \delta \widetilde{P}_{a_1}, \delta \widetilde{Q}_{a_2}, \delta \widetilde{P}_{a_2}, \delta \widetilde{Q}_{c_1}, \delta \widetilde{P}_{c_1}, \delta \widetilde{Q}_{c_2}, \delta \widetilde{P}_{c 2}\right)^T $ and 
$$\mathcal{Y}(t)=\left(\sqrt{\gamma} \widetilde{Q}_{a_1}^{i n}, \sqrt{\gamma} \widetilde{Y}_{a_1}^{i n}, \sqrt{\gamma} \widetilde{Q}_{a_2}^{i n}, \sqrt{\gamma} \widetilde{P}_{a_2}^{i n}, \sqrt{\Gamma} \widetilde{Q}_{c_1}^{i n}, \sqrt{\Gamma} \widetilde{P}_{c 1}^{i n}, \sqrt{\Gamma} \widetilde{Q}_{c_2}^{i n}, \sqrt{\Gamma} \widetilde{P}_{c 2}^{i n}\right)$$
where $\mathcal{Z}(t)$ is the quadrature vector and $\mathcal{Y}(t)$ is the noise vector, the matrix $\mathcal{Q}$ takes the following form
\begin{equation}
\mathcal{Q}=\left(\begin{array}{cccccccc}
-\gamma / 2 & 0 & 0 & -\beta & \mathcal{J} & 0 & 0 & 0 \\
0 & -\gamma / 2 & \beta & 0 & 0 & \mathcal{J} & 0 & 0 \\
0 & -\beta & -\gamma / 2 & 0 & 0 & 0 & \mathcal{J} & 0 \\
\beta & 0 & 0 & -\gamma / 2 & 0 & 0 & 0 & \mathcal{J} \\
-\mathcal{J} & 0 & 0 & 0 &-\frac{\Gamma}{2}+2 \lambda\cos (\theta)  & 2 \lambda \sin (\theta) & 0 & - \alpha \\
0 & -\mathcal{J} & 0 & 0 & 2 \lambda \sin (\theta) & -\frac{\Gamma}{2}-2 \lambda \cos (\theta) & \alpha & 0 \\
0 & 0 & - \mathcal{J} & 0 & 0 & -\alpha & -\frac{\Gamma}{2}+2 \lambda \cos (\theta) & 2 \lambda \sin (\theta) \\
0 & 0 & 0 & -\mathcal{J} & \alpha & 0 & 2 \lambda \sin (\theta) & -\frac{\Gamma}{2}-2 \lambda \cos (\theta)
\end{array}\right) .
\end{equation}
The eigenvalues of matrix $\mathcal{Q}$ has negative real parts  which makes the system stable with the experimental parameters reported in reference \cite{455}. This is in accordance with the Routh-Hurwitz criterion \cite{jamila}. We can describe the system in the steady state by the Lyapunov equation \cite{lpnv}
\begin{equation}
\mathcal{Q} \eta+\eta \mathcal{Q}^T=-\Omega
\end{equation} 
where $\Omega$ is the stationary noise matrix:  $\Omega_{i j} \delta\left(t-t^{\prime}\right)=(1 / 2)\left\langle \mathcal{Y}(t)_i^{i n}(t) \mathcal{Y}(t)_j^{i n}\left(t^{\prime}\right)+\mathcal{Y}(t)_j^{i n}\left(t^{\prime}\right) \mathcal{Y}(t)_i^{i n}(t)\right\rangle$. $\eta$ is the covariance matrix representing the system
$\eta_{i j}=(1 / 2)\left(\left\langle \mathcal{Z}_i(t) \mathcal{Z}_j\left(t^{\prime}\right)+\right.\right.$ $\left.\left.\mathcal{Z}_j\left(t^{\prime}\right) \mathcal{Z}_i(t)\right\rangle\right)$.\\
$\Omega$ take the following expression
\begin{equation} 
\Omega=\left(\begin{array}{cccccccc}\gamma^{\prime} & 0 & 0 & 0 & 0 & 0 & 0 & 0 \\ 0 & \gamma^{\prime} & 0 & 0 & 0 & 0 & 0 & 0 \\ 0 & 0 & \gamma^{\prime} & 0 & 0 & 0 & 0 & 0 \\ 0 & 0 & 0 & \gamma^{\prime} & 0 & 0 & 0 & 0 \\ 0 & 0 & 0 & 0 & \Gamma^{\prime} & 0 &\mathcal{V}  \Gamma & 0 \\ 0 & 0 & 0 & 0 & 0 & \Gamma^{\prime} & 0 & -\mathcal{V} \Gamma \\ 0 & 0 & 0 & 0 & \mathcal{V} \Gamma & 0 & \Gamma^{\prime} & 0 \\ 0 & 0 & 0 & 0 & 0 & -\mathcal{V} \Gamma & 0 & \Gamma^{\prime}\end{array}\right)
\end{equation}
where
$\gamma^{\prime}=\gamma\left(nth+\frac{1}{2}\right)$ and  $\Gamma^{\prime}=\Gamma\left(\mathcal{R}+\frac{1}{2}\right)$.	
\section{Quantum entanglement}
This section summarizes the definition of logarithmic negativity used in this study to quantify quantum entanglement between the two mechanical modes.
It is possible to evaluate the nonclassical correlations within the bipartite subsystem composed of moving mirrors $M1$ and $M2$ by using logarithmic negativity as a measure of entanglement.
It is possible to simplify the global covariance matrix of the two mechanical modes in the following matrix
\begin{equation}
\label{covar}
\sigma=\left(\begin{array}{cc}
X & Z \\
Z^T & Y
\end{array}\right)
\end{equation}
Let $X$ and $Y$ denote the covariance matrices of individual modes, each having dimensions $2\times 2$. The $2\times 2$ covariance matrix $Z$ describes the correlation between the two mechanical subsystems. In the context of continuous variable (CV) systems, the logarithmic negativity $E_{N}$ can be defined by \cite{34,35}
\begin{equation}
\label{34}
E_{N} = \max \left[ 0,- ln(2\varrho^{-})\right] 
\end{equation}
where $\varrho^{-}$ is the smallest symplectic eigenvalue measuring the entanglement between the two mechanical modes, defined as follows
\begin{equation}
\varrho^{-}=\bigg[\frac{\Psi-\sqrt{\Psi^2-4 \operatorname{det} \sigma}}{2}\bigg]^{1/2}
\end{equation}
$\Psi= \operatorname{det} X + \operatorname{det} Y- 2 \operatorname{det} Z $ is a function of elements of symplectic of the simplified covariance matrix $\sigma $.
The system is separable when $\varrho^{-} > \frac{1}{2}$.
\section{Results and discussion}	
We will discuss the steady-state quantum correlations of the two mechanical modes against various effects using the experimental values reported in  \cite{455}: each  movable mirror has a mas $m= 145$ $\mathrm{ng}$ oscillates with frequency $\omega_{m}= 2 \pi \times 947\times 10^{3} $ $\mathrm{Hz}$. The laser that drives the system possesses a power of $p=11$ $\mathrm{mW}$. The cavity length and frequency are, respectively $L= 25$ $\mathrm{mm}$ and $\omega_c=2 \pi \times 2.82\times 10^{14}$ $\mathrm{Hz}$, the laser frequency is $\omega_{l}= 2 \pi \times 5.26 \times 10^{14} $ $\mathrm{Hz}$,  $\Gamma= 2 \pi \times 215 \times 10^{3}$ $\mathrm{Hz}$ and $\gamma=2 \pi \times 140 \times 10^{3} $ $\mathrm{Hz}$.
\begin{figure}[h!]
\begin{center}
\fbox{ \includegraphics[width = 8cm,height=6cm]{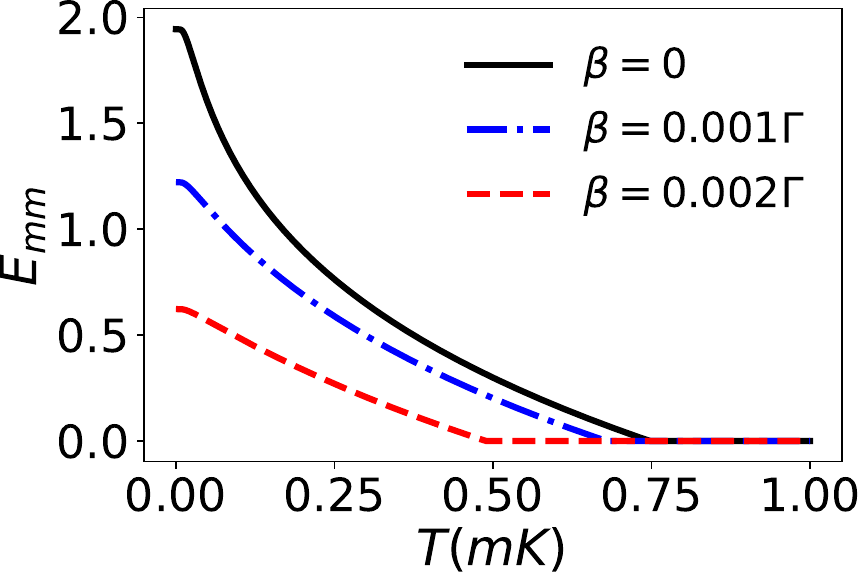} }
\caption{The logarithmic negativity $E_{mm}$ between the two mechanical modes as a function of temperature \textrm{T} for various values of the phonon tunneling rate $\beta$, with $r=1.5$, $\theta=0$, $\lambda=0.2\Gamma$ and $\alpha=0.0015\Gamma$.}\label{2}
\end{center}
\end{figure}
In Fig. \ref{2}, we display how entanglement between the two mechanical modes changes with temperature for various values of the coupling rate  $\beta$. One can observe that for a fixed value of $\beta $, the entanglement decreases as temperature increases due to decoherence phenomena \cite{dec}. The optimal value of $T$ beyond which the entanglement disappears, decreases with increasing $\beta$, and as the tunneling rate grows, the maximum value of entanglement decreases for a fixed temperature. This indicating that the tunneling rate degrades the entanglement as it mentioned in \cite{hmj}.
\begin{figure}[h!]
\begin{center}
\fbox{ \includegraphics[width = 8cm,height=6cm]{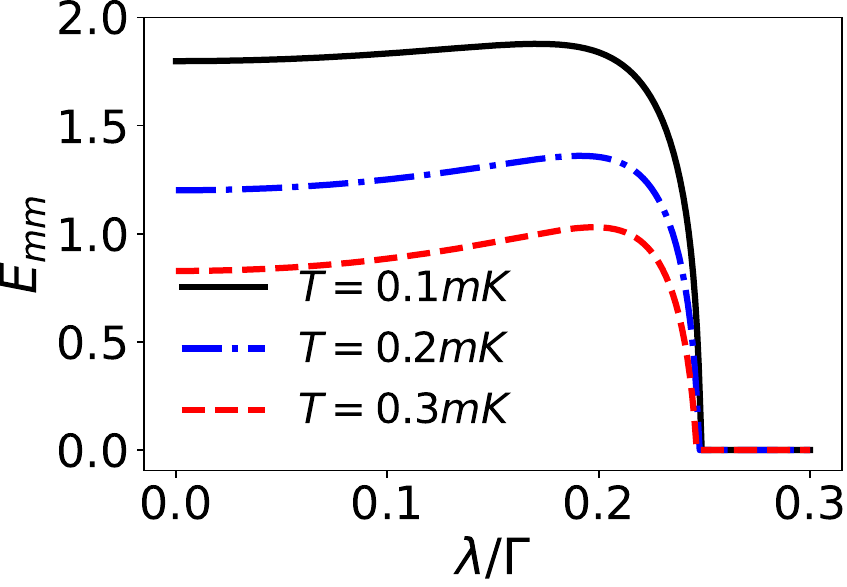}} 
\caption{The logarithmic negativity $E_{mm}$ between the two mechanical modes versus the gain $\lambda$ of the degenerate parametric amplifier for differents values of the temperature $T$, with $r=3$, $\theta=0$, $\beta=0.0002\Gamma$ and $\alpha=0.0015\Gamma$.}\label{6}
\end{center}
\end{figure}
 
We plot in Fig. \ref{6} the entanglement between the two mechanical oscillators $E_{mm}$ as a function of the gain $\lambda$ of the parametric amplifier, considering different temperature values and for a fixed values of all other parameters. For fixed values of $T$, we notice that entanglement is enhanced with increasing the values of $ \lambda $ until it reaches its maximum value $\lambda_{max}$, then, the entanglement decreases quickly with $\lambda $ increasing. The gain $\lambda $ is influenced by the number of photons present in the cavity. Thus, the increase in this number of photons leads to the emergence of thermal effects causing the phenomenon of decohernce \cite{48}. This phenomenon causes the decay of quantum correlations, a topic that has been previously discussed in \cite{k}.
\begin{figure}[h!]
\begin{center}
\fbox{ \includegraphics[width = 8cm,height=6cm]{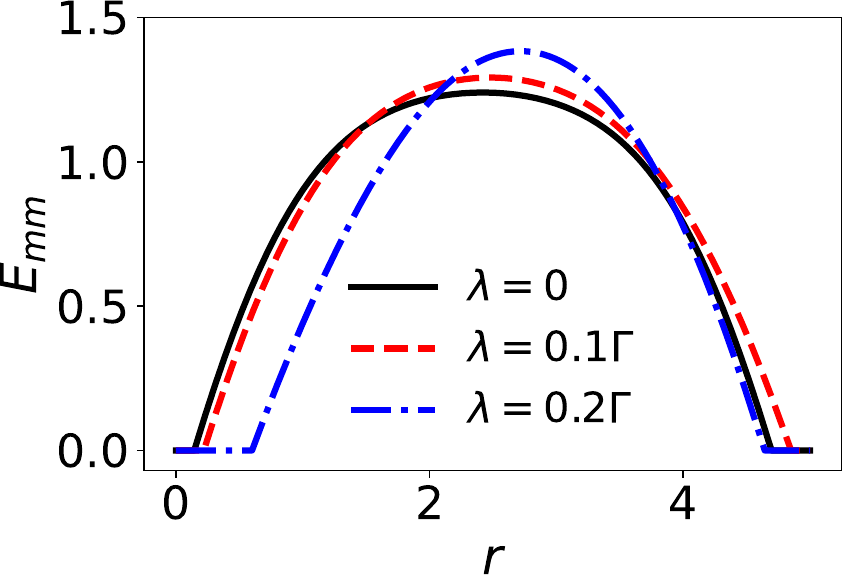}}
\caption{The logarithmic negativity $E_{mm}$ of two mechanical modes as a function of the parameter $r$ characterizing the squeezed light for various values of the gain $ \lambda$ of parametric amplifier, with $T=0.2$, $\theta=0$ mK, $\beta=0.0002\Gamma$ and $\alpha=0.0015\Gamma$.}
\label{rl}
\end{center}
\end{figure}

In Fig. \ref{rl}, we present a plot showing the entanglement $E_{mm}$ of mechanical modes as a function of the squeezing parameter $r$ for various values of the gain $ \lambda$ of the parametric amplifier. This figure reveals that the creation of entanglement between movable mirrors requires a minimum value $r_{min}$ of the parameter $r$, which is greater than zero ($r_{min}>0$). For a fixed value of $ \lambda$ the entanglement is achieved when $r$ exceeds $r_{min}$, and interestingly $r_{min}$ increases with higher values of $ \lambda$, a phenomenon known as entanglement sudden birth \cite{sd}. Additionally, we observe that for a given value of $ \lambda $, the amount of entanglement increases as $r$ increases until it reaches its maximum value, this occurrence can be attributed to the resonance phenomenon between the two movable mirrors \cite{mm}. However after reaching the maximum value, the entanglement decreases rapidly, this decrease, even when $r$ increases, can be attributed to the diminishing effect of radiation pressure \cite{mm,31}. Lastly, the plot in Fig. \ref{rl} demonstrates that the entanglement is affected by the value of the gain of the parametric amplifier . As expected, there exists a direct relationship between the generation of entanglement among mechanical modes, the gain of the parametric amplifier  and the presence of squeezed light.
\begin{figure}[h!]
\begin{center}
\fbox{ 
\includegraphics[width = 8cm,height=6cm]{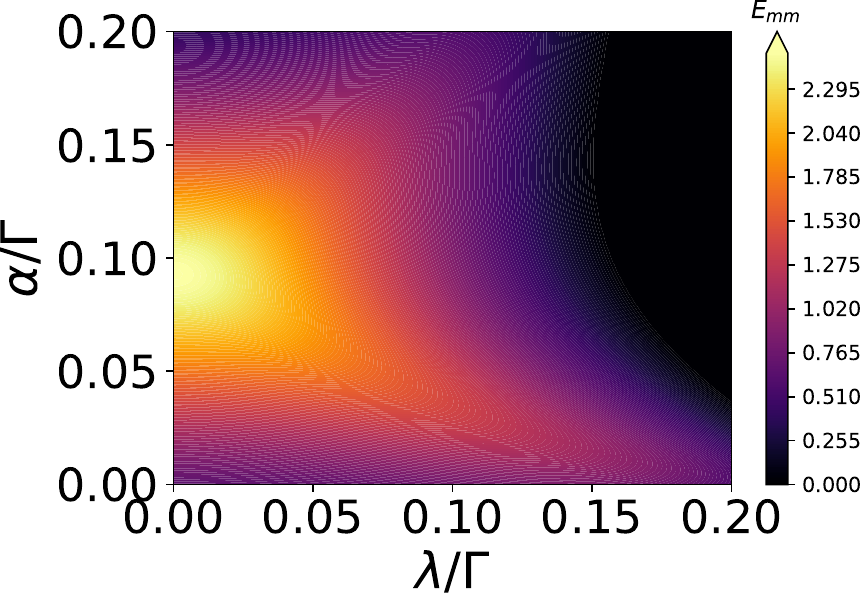}}
\caption{The density plot of bipartite bipartite entanglement between the two mechanical resonators as a function of coupling strength $\alpha $ of the photon hopping (PH) and the gain $\lambda$ of parametric amplifier. With $r=2$, $\theta=0$, $T=0.02$ mK and $\beta=0.002\Gamma$.}.
\label{5}
\end{center}
\end{figure}

We plot in Fig. \ref{5}, the entanglement $E_{mm}$ between the two mechanical resonators as a function of coupling strength $\alpha$ and the gain $\lambda$. We remark that the entanglement decreases with increasing $ \lambda$ as already mentioned in Fig. \ref{6}. The magnitude of the stationary entanglement between the two mechanical modes improved only for a specific region of the $\alpha $ values. This point highlights the significance of selecting an appropriate value of  $\alpha$ to enhance the quantum entanglement, as shown in \cite{31}.\\
For a value of $ \alpha $ almost larger than $0.05\Gamma$, we observe that the entanglement decreases with the increase of $\lambda$, until the sudden death of the entanglement \cite{SOL4}. 	
\section{Conclusion}
In conclusion, we have proposed a schematic of an hybrid optomechanical system to discuss the enhancement of entanglement between two movable mirrors (mechanical modes) in a double-cavity optomechanical system via squeezed vacuum injection and intracavity squeezed light. The optical modes interact with each other via photon hopping, while the mechanical resonators are coupled through phonon tunneling. The two cavities are pumped by squeezed light and driven by coherent laser sources. We have proposed the logarithmic negativity as a quantum measure that can be used in this proposed system. By adjusting the gain of the parametric amplifier, we successfully increased the entanglement for specified values of $\lambda$. Also, we showed that the entanglement between the two movable mirrors is influenced by the rate of phonon tunneling; when the rate $\beta$ increases, the entanglement decreases, indicating that the coupling rate of phonons degrades entanglement. In addition, entanglement can be improved by a convenient choice of coupling strength in the case of the photon hopping process. We have particularly witnessed two phenomena: entanglement in sudden birth and entanglement in sudden death.

\end{document}